\documentclass{elsart}
\usepackage{epsfig}
\usepackage{amssymb}
\usepackage{graphics}
\usepackage[dvips]{color}
\newcommand{\nn}{\nonumber}

\newcommand{\be}{\begin{equation}}
\newcommand{\ee}{\end{equation}}
\newcommand{\bea}{\begin{eqnarray}}
\newcommand{\eea}{\end{eqnarray}}

\def\lsim{\buildrel < \over {_{\sim}}}
\def\gsim{\buildrel > \over {_{\sim}}}

\def\simge{\mathrel{%
   \rlap{\raise 0.511ex \hbox{$>$}}{\lower 0.511ex \hbox{$\sim$}}}}
\def\simle{\mathrel{
   \rlap{\raise 0.511ex \hbox{$<$}}{\lower 0.511ex \hbox{$\sim$}}}}
\begin{document}

\begin{frontmatter}


\begin{flushright}
%{\tt hep-ph/ }\\
{ROMA-1438-06}
\end{flushright}
\title{Total neutrino and antineutrino nuclear cross sections around 
1 GeV}

\author[INFN,Dipartimento]{Omar Benhar}
\author[INFN,Dipartimento]{, Davide Meloni}
\address[INFN]{INFN, Sezione di Roma, I-00185 Roma, Italy}
\address[Dipartimento]{Dipartimento di Fisica, Universit\`a ``La Sapienza'', 
I-00185 Roma, Italy}

%\author{}

%\address{}

\begin{abstract}
% Text of abstract
We investigate neutrino-nucleus interactions at
energies around 1 GeV. In this regime, the main contributions to the
cross sections come from quasi-elastic and $\Delta$ production processes.
Our formalism, based on the Impulse Approximation is well suited to describe both
types of interactions. We focus on a series of important nuclear effects in 
the interaction of electron neutrinos with $^{16}O$, also relevant to future 
$\beta$-Beams.
Our results show that the Fermi gas model, widely used in data analysis
of neutrino experiments, overestimates the total cross sections by as 
much as 20 \%.
\end{abstract}

\end{frontmatter}

% main text
%%%%%%%%%%%%%%%%%%%%%%%%%%%%%%%%%%%%%%%%%%%%%%%%%%%%%%%%%%%%%%%%%%%%%%%%%%%%%%%%%%%%%%%
\section{Introduction}
\label{intro}
The field of neutrino physics is rapidly developing after atmospheric and solar
neutrino oscillations have been established.
The results of atmospheric, solar, accelerator and reactor neutrino experiments \cite{exp} show 
that flavour mixing occurs not only in the hadronic sector, as it has been long known, but 
in the leptonic sector as well. The experimental results point to two very distinct mass 
differences\footnote{A third mass difference, $\Delta m^2_{LSND} \sim 1$ eV$^2$, suggested by 
the LSND experiment \cite{lsnd}, has not been confirmed yet \cite{boone}.}, 
$\Delta m^2_{sol} \approx 8.2 \times 10^{-5}$ eV$^2$ and 
$|\Delta m^2_{atm}| \approx 2.5 \times 10^{-3}$ eV$^2$.
Only two out of the four parameters of the three-family leptonic mixing matrix $U_{PMNS}$ 
\cite{neutrino_osc} are known: $\theta_{12} \approx 34^\circ$ and $\theta_{23}\approx 45^\circ$. 
The other two parameters, $\theta_{13}$ and $\delta$, are still unknown: for the mixing angle
$\theta_{13}$ direct searches at reactors \cite{chooz} and three-family global analysis of the 
experimental data \cite{globalfit,Gonzalez-Garcia:2004jd} give the upper bound $\theta_{13} 
\leq 11.5^\circ$, whereas for the leptonic CP-violating phase $\delta$ we have no information 
whatsoever. Two additional discrete unknowns are the sign of the atmospheric mass difference and 
the $\theta_{23}$-octant (if $\theta_{23} \neq 45^\circ$).

The full understanding of the leptonic mixing matrix, together with the discrimination 
of the Dirac/Majorana character of neutrinos and with the measurement of their 
absolute mass scale, represents the main goal of neutrino physics in the next decade. 
The SK detector has gathered indirect evidence of $\nu_\mu \to \nu_\tau$ 
conversion of atmospheric neutrinos whereas 
the SNO detector \cite{SNO} has shown that a fraction of the $\nu_e$'s emitted by the Sun core reaches 
the
Earth-located detectors converted into $\nu_\mu$'s and $\nu_\tau$'s (and not into unobservable sterile neutrinos). 
On the other hand, new-generation experiments have been proposed to look for the fleeting 
and intimately related parameters $\theta_{13}$ and $\delta$ through the more promising 
``appearance channels'' such as $\nu_e \to \nu_\mu$  
(the ``golden channel'' \cite{Cervera:2000kp}) 
and $\nu_e \to \nu_\tau$ (the ``silver channel'' \cite{silver}). However, strong correlations between $\theta_{13}$ and 
$\delta$ and the presence of parametric degeneracies in the 
($\theta_{13},\delta$) parameter space \cite{Burguet-Castell:2001ez,Barger:2001yr}, make 
the simultaneous measurement of the two variables extremely difficult. 
A further problem arises from our present imprecise knowledge of atmospheric parameters, 
whose uncertainties
are far too large to be neglected when looking for such tiny signals as those expected in appearance 
experiments 
%driven by the $\nu_\mu \to \nu_e$ and $\nu_e \to \nu_\mu, \nu_\tau$ oscillation probabilities 
\cite{Donini:2005rn}.
Most of the proposed solutions to avoid all these problems suggest
to combine different experiments and facilities, such as
Super-Beams (of which T2K \cite{Itow:2001ee} is the first approved one), 
$\beta$-Beams \cite{Zucchelli:sa} 
or Neutrino Factories \cite{Geer:1997iz}, although it is not completely clear whether
the sensitivity of such experiments will be enough.

In view of these developments, it is vital to reduce as much as possible
any source of uncertainty that could ruin the required precision
and, among them, the knowledge of the
neutrino-nucleus cross sections is one of the thorniest problems to be faced. In
fact, the data are either very few (the case of neutrinos) 
or missing altogether (the case of antineutrinos). On top of that, the few available 
data have generally not been taken on 
the same targets used in the experiments (either water, iron, lead or plastics), and the 
extrapolation from different nuclei 
is complicated by nuclear effects, that can play an important role at the 
considered energies. Among the proposed experiments, Miner$\nu$a 
\cite{manly} and SciBooNE \cite{tanaka} will have the capability for precise
measurements in a wide range of energies and different nuclear targets. 

\begin{figure}[t!]
\vspace{-0.5cm}
\begin{center}
\begin{tabular}{c}
\hspace{-0.3cm} \epsfxsize9cm\epsffile{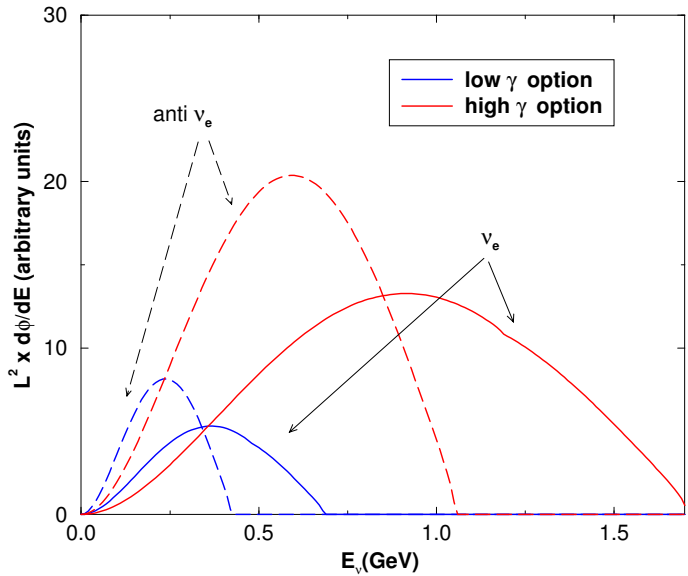}\\
\end{tabular}
\caption{\it \label{fig:fluxes}
$\beta$-Beam fluxes  as a function of neutrino energy for two specific sets of $\gamma$ 
values. Low and high $\gamma$ correspond to $(\gamma_{\nu_e},\gamma_{\bar
\nu_e})=(100,60)$ \protect\cite{Bouchez:2003fy}, and $(\gamma_{\nu_e},\gamma_{\bar
\nu_e})=(250,150)$ \protect\cite{Burguet-Castell:2003vv}, respectively.}
\end{center}
\end{figure}

In this paper we will focus on total nuclear cross sections at energies up to 
$\sim$ 1 GeV, relevant to a number of neutrino experiments as well as to the 
planned Super-Beams and $\beta$-Beams. As an example of the latter, 
in Fig.~\ref{fig:fluxes}, we show the energy spectra for two different setups, in which 
the three decay modes of $^{18}$Ne \cite{Donini:2004hu} are considered.

%%%%%%%%%%%%%%%%%%%%%%%%%%%%%%%%%%%%%%%%%%%%%%%%%%%%%%%%%%%%%%%%%%%%%%%%%%%%%%%%%%%%%%%%%%%%%
%%%%%%%%%%%%%%%%%%%%%%%%%%%%%%%%%%%%%%%%%%%%%%%%%%%%%%%%%%%%%%%%%%%%%%%%%%%%%%%%%%%%%%%%%%%%%

%The mean energy of the \nubare, \nue\ beams for the {\it low-energy option} 
%are 0.25~GeV and 0.4~GeV, respectively, whereas the higher $\gamma$ scenario
%favours 0.6~GeV and 0.9~GeV with neutrino energies extending up to 2~GeV.
%This is the energy regime we will deal with in the rest of the paper. 

In this energy regime, the dominant contribution to the charged lepton production
cross section comes from quasi-elastic reactions. However, for 
$E_\nu~\simge$~0.5~GeV, inelastic production of charged leptons through
excitation of the $\Delta$ (as well as other) resonance becomes relevant. 
At higher energies,
deep inelastic scattering also becomes progressively important, but the impact on the
energy regime under investigation is expected to be negligible. Therefore, we do not
take this contribution into account.

Quasi-elastic scattering, pion production and deep inelastic scattering are combined
in event generators (e.g. NUANCE \cite{casper}, NEUGEN \cite{gallagher}
and NEUT \cite{hayato}), that model the neutrino detector response using a
simplified picture, referred to as Relativistic Fermi Gas Model (RFGM), 
in which the nucleus is described as a collections of quasi-free nucleons. 
A wealth of high-precision electron scattering data have shown that the RFGM neglects
important features of nuclear dynamics \cite{book}. The most important is the presence of strong 
nucleon-nucleon correlations, leading to the appearance of high momentum and high energy
components in the energy-momentum distribution of the nucleons in the target 
nucleus \cite{BP_RMP}.

In this paper we go beyond the simple RFGM, and 
analyse neutrino-nucleus interactions at intermediate energies
using a more realistic description of nuclear dynamics, based on nuclear many-body 
theory (NMBT). 
We evaluate the cross sections for inclusive processes, in which only the outgoing 
charged lepton is detected, including both 
quasi-elastic and pion production processes.
Our formalism is based on the impulse approximation (IA) picture, which is expected 
to be applicable at the large momentum transfers corresponding to beam energies around 
1 GeV. Final state interactions (FSI)
between the hadrons produced at the elementary weak interaction vertex and the 
spectator nucleons have been not taken into account, as they do not contribute
to the total cross section. However, the effect of statistical correlations between the struck 
nucleon and the spectators, which are known to be important in quasi-elastic scattering at 
low $Q^2$, has been included.

The plan of the paper is the following.
In Section~\ref{IA} we recall the formalism of the IA used to describe
charged current (CC) interactions. In Section~\ref{elastic} we
show our results for the inclusive CC cross sections, while in
 Section~\ref{resonance} we present the results for resonance production. 
Finally we draw the conclusions and outline the future prospects of our work. The
appendices contain formulae that, together with those presented in the main
text, will enable the readers to write their own codes
to compute the cross sections discussed in this paper.
%%%%%%%%%%%%%%%%%%%%%%%%%%%%%%%%%%%%%%%%%%%%%%%%%%%%%%%%%%%%%%%%%%%%%%%%%%%%%%%%%%%%%%%
\section{The Impulse Approximation}
\label{IA}

\begin{figure}[b!]
\begin{center}
\begin{tabular}{c}
\hspace{-0.3cm} \epsfxsize6cm\epsffile{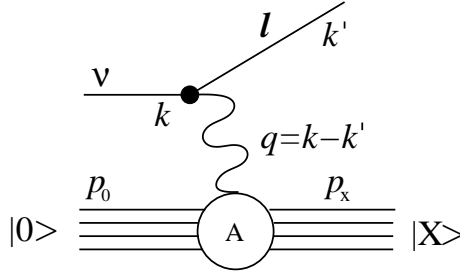}\\
\end{tabular}
\caption{\it \label{int}
Feynman diagram for the process $\nu_\ell + A \to \ell^- + X$.}
\end{center}
\end{figure}
The differential cross section for the process
$\nu_\ell + A \to \ell^- + X$ (Fig.~\ref{int}) in which a neutrino carrying initial
four-momentum $k=(E_\nu,\bf k)$ scatters off a nuclear target to a state of four-momentum
$k^{'}=(E_\ell,\bf k^{'})$, the target final state being undetected, can be written
in Born approximation:
\bea
\label{elem}
\frac{d^2\sigma}{d\Omega dE_\ell}=\frac{G_F^2\,V^2_{ud}}{16\,\pi^2}\,
\frac{|\bf k^{'}|}{|\bf k|}\,L_{\mu\nu}\, W_A^{\mu\nu} \ , 
\eea
where $G_F$ is the Fermi constant and $V_{ud}$ is the CKM matrix element coupling $u$
and $d$ quarks. The leptonic tensor, that can be written in the form 
\bea
\label{leptensor}
L_{\mu\nu}&=&8\,\left[k_\mu^{'}\,k_\nu+k_\nu^{'}\,k_\mu- g_{\mu\nu}(k\cdot
k^{'})-i\,\varepsilon_{\mu\nu\alpha\beta}\,k^{'\beta}\,k^\alpha \right]
\eea
is completely determined by lepton kinematics, whereas the nuclear tensor 
$W_A^{\mu\nu}$, containing all the information on strong interactions dynamics,
describes the response of the target nucleus. Its definition
involves the initial and final hadronic states $|0\rangle$ and $|X\rangle$,
carrying four momenta $p_0$ and $p_X$, respectively, as well as the nuclear
electroweak current operator $J^\mu_A$:
\bea
\label{hadronictensor}
W_A^{\mu\nu}&=& \sum_X \,\langle 0 | {J_A^\mu}^\dagger | X \rangle \,
      \langle X | J_A^\nu | 0 \rangle \;\delta^{(4)}(p_0 + q - p_X) \ , 
\eea
where the sum includes all hadronic final states.
Calculation of $W_A^{\mu\nu}$ at moderate momentum trasfers ($|\bf q| < $0.5 GeV)
can be carried out within NMBT using nonrelativistic wave
functions to describe the initial and final states and expanding the current
operator in powers of ${\bf q}/m_N$, $m_N$ being the nucleon mass. However, at
higher values of $|\bf q|$, corresponding to $E_\nu \gsim 1$ GeV, 
we can no longer describe the final states $|X\rangle$ in terms of nonrelativistic
nucleons. Calculation of $W_A^{\mu\nu}$ in this regime requires a set
of simplifying assumptions, allowing one to take into account the relativistic
motion of final state particles carring momentum $\sim {\bf q}$, as well as the occurrence
of inelastic processes, leading to the appearance of hadrons other than protons
and neutrons.

\begin{figure}[h!]
\vspace{-0.5cm}
\begin{center}
\begin{tabular}{c}
\hspace{-0.3cm} \epsfxsize7cm\epsffile{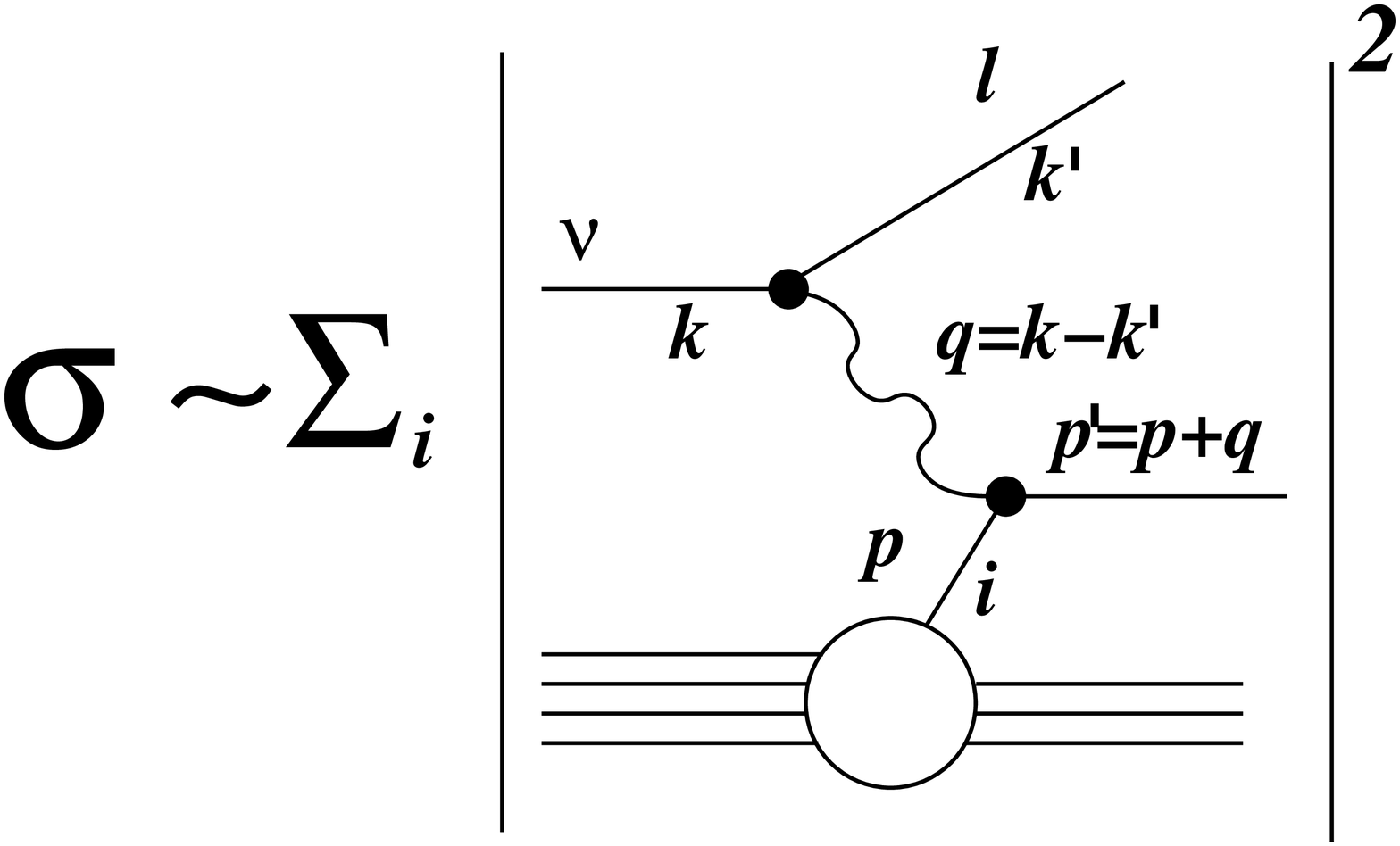}\\
\end{tabular}
\caption{\it \label{int2}
Diagrammatic representation of the process $\nu_\ell + A \to \ell^- + X$ in the 
Impulse Approximation scheme.}
\end{center}
\end{figure}

In the rest of the paper, we adopt the IA scheme, based on the
assumptions that at large enough ${\bf q}$ the target nucleus is seen by the
probe as a collection of individual nucleons and that the particles produced at
the interaction vertex and the recoiling $(A-1)$-nucleon system evolve
independently (see Fig.~\ref{int2} for a pictorial representation of the IA
picture). As a consequence, IA neglects both statistical correlations due to Pauli
blocking and the rescattering processes driven by strong interactions (FSI).

Within this picture, the nuclear current can be written as a sum of one-body
currents, i.e. $J^\mu_A \rightarrow  \sum_i \, J^\mu_i$, while the final state reduces to the direct product
of the hadronic state produced at the weak vertex (with momentum ${\bf p^{'}}$)
and that describing the $(A-1)$-nucleon residual system, with momentum $\bf p_{\cal R}$:
$| X \rangle \to | i,{\bf p}^{'} \rangle \otimes | {\cal R}, \bf p_{\cal R} \rangle$.
The calculation of the weak tensor is described in Ref. \cite{Benhar:2005dj}.
Here we only quote the final result:

\bea
W_A^{\mu\nu}&=& \frac{1}{2}\int d^3p\,dE \,P({\bf
p},E)\frac{1}{4\,E_{|\bf p|}\,E_{|\bf p+q|}} \,  W^{\mu\nu}(\tilde p,\tilde
q) \ ,
\label{hadtensor}
\eea
where $E_{\bf p}=\sqrt{|{\bf p}|^2+m_N^2}$. The above definition deserves some
comments. The function $P({\bf p},E)$ is the target {\it spectral function}, i.e.
 the probability distribution of finding a nucleon with momentum 
${\bf p}$ and removal energy $E$ in the target nucleus. It then encodes all the 
informations about the
initial (struck) particle. The quantity $W^{\mu\nu}$ is the tensor describing
the weak interactions of the $i$-th nucleon in free space; the effect of
nuclear binding of the struck nucleon is accounted for by the
replacement:
\bea
\nn
q=(\nu,{\bf q}) \to \tilde q=(\tilde \nu,{\bf q})
\eea
with $\tilde \nu = E_{|\bf p+q|}-E_{|\bf p|}$. In particular, the variable
$\tilde \nu$ takes into account the fact that a fraction $\delta \nu$ of the energy transfer
goes into excitation energy of the spectator system, and that the elementary
scattering process can be described as if it took place in free space with
energy transfer $\tilde \nu = \nu - \delta\nu$.
It follows that the second argument in the hadronic tensor is $\tilde p=(E_{|\bf p|},\bf p)$.
Substituting Eq.~(\ref{hadtensor}) into Eq.~(\ref{elem}), we get the final formula
for the {\it nuclear} cross section:
\bea
\label{final_cs}
\frac{d^2\sigma_{IA}}{d\Omega dE_\ell}=
\int d^3p\,dE \,P({\bf
p},E)\,\frac{d^2\sigma_{\rm elem}}{d\Omega dE_\ell} \ ,
\eea
in which we have redefined the {\it elementary} cross section as 
\bea
\label{elem_cs}
\frac{d^2\sigma_{\rm elem}}{d\Omega dE_\ell}=\frac{G_F^2\,V^2_{ud}}{32\,\pi^2}\,
\frac{|\bf k^{'}|}{|\bf k|}\,\frac{1}{4\,E_{\bf p}\,E_{|\bf p+q|}}\,L_{\mu\nu}
W^{\mu\nu} \ .
\eea
The integration limits for Eq.~(\ref{final_cs}) are given in Appendix ~\ref{int_limit}.

The hadronic tensor $W_{\mu\nu}$ can be written in terms of five structure functions
$W_i$ as 
\bea
\label{hadrdec}
W^{\mu\nu}&=&-g^{\mu\nu}\,W_1+\tilde p^\mu\,\tilde p^\nu\,\frac{W_2}{m_N^2}+
i\,\varepsilon^{\mu\nu\alpha\beta}\,\tilde q_{\alpha}\,\tilde p_\beta\,
\frac{W_3}{m_N^2}+\tilde q^\mu\,\tilde q^\nu\,\frac{W_4}{m_N^2}+\\
&&(\tilde p^\mu\,\tilde q^\nu+\tilde p^\nu\,\tilde q^\mu)\,\frac{W_5}{m_N^2}
\nn \ .
\eea
Its contraction with the leptonic tensor (\ref{leptensor}) yields
\bea
\label{contraction}
L^{\mu\nu}\,W_{\mu\nu}&=&16\,\sum_i \,W_i\,\left(\frac{A_i}{m_N^2}\right) \ .
\eea

The functions $A_i$, containing kinematical factors, are collected in
Appendix ~\ref{a_coeff} whereas the structure functions $W_i$, expressed in 
terms of form factors, are given in Appendix~\ref{formfactors}.

Eq.~(\ref{final_cs}) is a general formula that can be used to
describe quasi-elastic scattering and resonance production in neutrino-nucleus charged
current interactions. The explicit expressions differ in the analytical form 
of the relevant form factors and for the replacement of the energy conserving 
$\delta$-function with a Breit-Wigner factor accounting for the finite width of the 
resonance.

%%%%%%%%%%%%%%%%%%%%%%%%%%%%%%%%%%%%%%%%%%%%%%%%%%%%%%%%%%%%%%%%%%%%%%%%%%%%%%%%%%%%%%%
\section{Elastic cross sections}
\label{elastic}
To obtain the elastic cross sections from Eq.~(\ref{final_cs}) we need to
clarify the form of
the structure functions $W_i$ and the meaning of the spectral function. The first part of the
job is easily done once we express the structure functions in terms of form 
factors $F_1,F_2,F_A,F_P$ using the 
hypotheses of CVC (Conserved Vector Current) and PCAC (Partially Conserved Axial Current), as
usually done for scattering off a free nucleon. 
In addition, the definition of the structure functions includes an energy
conserving $\delta$-function $\delta(s-m_N^2)$, where $s=(\tilde p+\tilde q)^2$ is the squared invariant 
mass of the final state of the struck nucleon.
In Appendix~\ref{formfactors} we collect the expression of 
$W_i$ in terms of form factors. Note that, except where differently
stated,  we use the values 
$M_V=0.84$ GeV and $M_A=1.05$ GeV for the vector and axial mass, respectively.

\subsection{Fermi gas and spectral function}
The RFGM \cite{Smith:1972xh}, widely used in Monte Carlo simulations, provides the 
simplest form of the spectral function: 
\bea
\label{fermigas}
P_{RFGM}({\bf p},E)=\left(\frac{6\,\pi^2\,A}{p_F^3}\right)\,\theta(p_F-{\bf p})\,
\delta(E_{\bf p}-E_B+E) \ ,
\eea
where $p_F$ is the Fermi momentum and $E_B$ is the average binding energy,
introduced to account for nuclear binding. The term in parenthesis is a constant
needed to normalize the spectral function to the number of target nucleons, $A$.
Thus, in this model $p_F$ and $E_B$ are two parameters that are {\it adjusted} to 
reproduce the experimental data. For oxygen, the analysis of electron scattering 
data yields $p_F=225$ MeV and $E_B=25$ MeV \cite{Moniz:1971mt}.
In more refined versions of RFGM, the binding energy
of the nucleon is not a constant but a momentum-dependent
function $V({|\bf p|})$ \cite{Brieva:1977dv}, entering the $\delta$-function in
Eq.~(\ref{fermigas}) according to $\delta(E_{\bf p}+V({|\bf p|})+E)$. 

Electron scattering data have provided overwhelming evidence that 
the energy-momentum distribution of nucleons in the nucleus is quite
different from that predicted by RFGM (see, e.g. Ref. \cite{Benhar_NUINT04}). 
The most important feature emerged from these data is the presence of strong
nucleon-nucleon (NN) correlations. 

Strong dynamical correlations give rise to virtual scattering processes leading to 
the excitation of the participating
nucleons to states of energy larger than the Fermi energy, thus depleting
the single particle levels within the Fermi sea. As a consequence, the spectral function
associated with nucleons belonging to correlated pairs extends to 
the region of $|{\bf p}| \gg p_F$ {\it and} $E \gg E_B$. 

Clearcut evidence of correlation effects has been provided by electron- and 
hadron-induced nucleon knock-out experiments \cite{specfact}. While the 
spectroscopic lines corresponding to knock out from the shell model states are
clearly seen in the measured energy spectra, the associated strengths are 
consistently lower that expected, regardless of the nuclear mass number.
Unfortunately, systematic measurements of the missing strength (typically about 20\%) 
pushed to higher energy by NN correlations, have not been carried out yet. However,
the results of a pioneering JLab experiment \cite{rohe2004} appear to be consistent 
with the expectations based on lower energy data.

Correlation effects can be consistently taken into account within NMBT, in which 
nuclei are described in terms of nonrelativistic nucleons interacting through 
the hamiltonian
\be
H_A = \sum_{i=1}^{A} \frac{{\bf p}_i^2}{2m} + \sum_{j>i=1}^{A} v_{ij}
 + ... \ ,
\label{H:A}
\ee
where $v_{ij}$ provides a quantitative account of the properties of the
two-nucleon system, i.e. deuteron properties and $\sim$ 4000 accurately measured
nucleon-nucleon scattering phase shifts at energies up to the pion production
threshold \cite{WSS}, while the ellipses refer
to a small additional three-nucleon potential. The energies
of the ground and low-lying excited states of nuclei with mass number
$A\leq10$ calculated within NMBT are in excellent agreement 
with the experimental values \cite{WP}. 

The calculation of $P({\bf p},E)$ within NMBT involves 
a degree of complexity that rapidly increases with $A$, so that it has been only 
carried out for $A\leq4$ \cite{dieperink,cps,sauer,ciofi4,morita,bp}.
However, thanks to the simplifications associated with translation invariance, 
highly accurate results are also available for uniform nuclear matter, 
i.e. in the limit A $\rightarrow \infty$ with Z=A/2 \cite{bff,pkebbg} ($Z$ denotes
the number of protons).

The spectral functions for medium-heavy nuclei, ranging from Carbon ($A=12$) to 
Gold ($A=197$), have been modeled using the
Local Density Approximation (LDA) \cite{bffs}, in which the experimental
information obtained from nucleon knock-out measurements is combined
with the results of theoretical calculations of the nuclear matter
$P({\bf p},E)$ at different densities. Within this approach,
$P({\bf p},E)$ is divided in two parts, corresponding to low momentum nucleons,
occupying shell model states, and high momentum nucleons, respectively.

The correlation contribution to $P({\bf p},E)$ of uniform nuclear matter 
has been calculated by Benhar {\it et al.} for a wide range of density values \cite{bffs}.
Within the LDA scheme, the results of Ref. \cite{bffs} can be used to obtain the 
corresponding quantity for a finite nucleus of mass number $A$. The full LDA 
nuclear spectral function can then be written as
\be
P_{LDA}({\bf p},E) = P_{MF}({\bf p},E) + P_{corr}({\bf p},E) \ .
\label{P:LDA}
\ee
From the spectral function, we can obtain the nucleon momentum 
distribution, defined as
\be
\label{def1:nk}
n({\bf p})  =  \int dE\ P({\bf p},E) \ .
\ee

\begin{figure}[h!]
\begin{center}
\epsfxsize10cm\epsffile{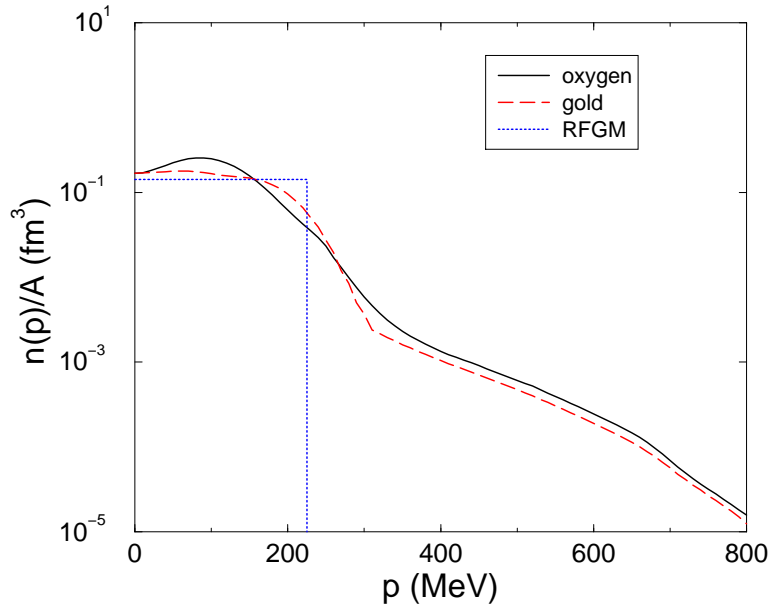}
\caption{\label{nk:O}\it
Momentum distribution of nucleons in the oxygen (solid line) and 
gold (dotted line) ground states.
Dashed line: RFGM with Fermi momentum $p_F = 225$ MeV and $E_B=25$ MeV.}
\end{center}
\end{figure}

The nucleon momentum distributions
of $^{16}$O and $^{197}$Au obtained from LDA, normalized to unity, 
are shown in Fig. \ref{nk:O}. 
For reference, the RFGM 
momentum distribution corresponding to Fermi momentum $p_F$ = 225 MeV is also 
shown by the dashed line. 

The results of Fig. \ref{nk:O}
show that $n({\bf p})/A$ becomes nearly independent of
$A$ at large $|{\bf p}|$ ($\gsim 300$ MeV/c). This feature
suggests that the correlation part of the spectral function
also scale with the target mass number, so that nuclear matter results
can be used at finite $A$.
It also clearly appears that the LDA $n({\bf p})$ is very different from 
that obtained from the RFGM, in which the large momentum tail is missing.

As pointed out in Section \ref{intro}, within IA all FSI are neglected.
The validity of this approximation has been thoroughly analyzed within the framework 
of non relativistic models of the nuclear response using correlated basis function (CBF) 
perturbation theory \cite{bff}.

In quasi-elastic inclusive processes {\it dynamical} FSI, which
are long known to be important in semi-inclusive and exclusive electron-nucleus
scattering processes, are much weaker. Their main effects are i) an energy shift of the 
differential cross section, due to the fact that the struck nucleon feels the mean field 
generated by the spectator particles and ii) a redistribution of the strength, leading 
to the quenching of the peak and the enhancement of the tails. 

Within non relativistic NMBT, it can be shown that  
FSI {\it do not} affect the {\it total} inclusive cross 
section, resulting from integration over the lepton variables 
(energy loss and scattering angle). This result follows from the fact that
the final states of the A-body system can be generated from the complete
set of asymptotic states, corresponding to independent propagation of the
struck nucleon and the recoiling spectator system, through a unitary
transformation. When the energy of the knocked out nucleon is large and
relativistic effects cannot be neglected, this procedure can still be
applied using a generalization of Glauber theory, based on the eikonal
approximation, as discussed in, e.g., Ref. \cite{Ben_Nik}. 
As our work is focused on total cross sections only, FSI effects will 
not be taken into account.

The effect of {\it statistical} correlations, arising from the antisymmetrization of 
the struck nucleon with respect to the spectator particles, is also disregarded within IA.
The results of calculations carried out within nonrelativistic models \cite{bff} show that 
Pauli blocking has negligible effect on the differential cross section for three-momentum transfer 
$|{\bf q}|$ larger than $\sim$ 500 MeV. However, as in the kinematical regime discussed in our 
work the total cross section includes significant contributions from scattering at lower $|{\bf q}|$,
the modification of the phase space available to the knocked-out nucleon must be taken into account.
A rather crude prescription to estimate the effect of Pauli blocking is based on the 
replacement \cite{Benhar:2005dj}
\be
P({\bf p},E) \rightarrow P({\bf p},E)
\theta(|{\bf p} + {\bf q}| - {\overline p}_F)
\label{pauli}
\ee
where ${\overline p}_F$ is the average nuclear Fermi momentum, defined as
\be
{\overline p}_F = \int  d^3r\ \rho_A({\bf r}) p_F({\bf r}),
\label{local:kF}
\ee
with $p_F({\bf r})=(3 \pi^2 \rho_A({\bf r})/2 )^{1/3}$, $\rho_A({\bf r})$ being the
nuclear density distribution. For oxygen, Eq. (\ref{local:kF}) yields
${\overline p}_F = 209$ MeV. Note that, unlike the spectral function, the
quantity defined in Eq. (\ref{pauli})
does not describe {\it intrinsic} properties of the target only, as it depends
explicitely on the momentum transfer. 
Overall, the approach described above involves no
adjustable parameters (as compared with other models discussed in Section \ref{comp}).

\begin{figure}[h!]
\begin{center}
\epsfxsize10cm\epsffile{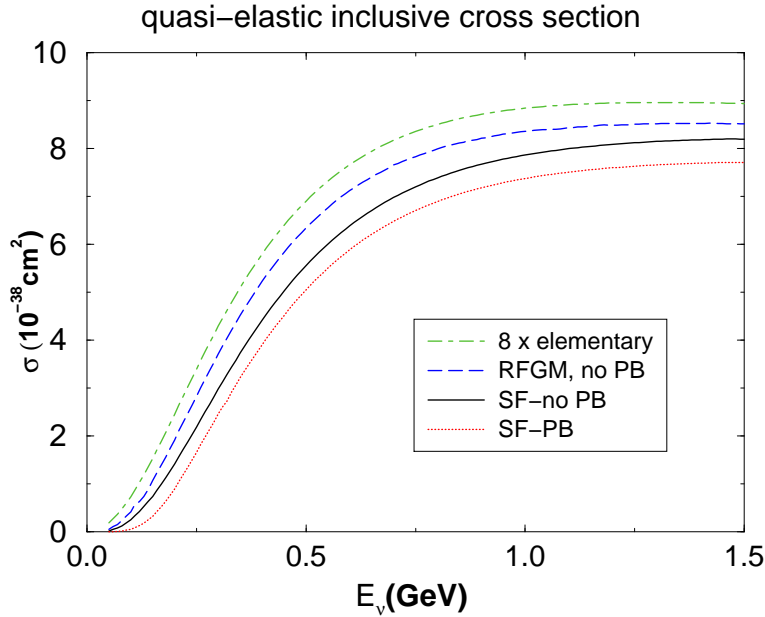}
\caption{\label{total_elastic}\it
Total quasi-elastic cross section for the process $\nu_e\,^{16}O \to e^-\,X$.
The dot-dash line represents eight times the elementary cross section; the dashed line is the 
result of the RFGM with Fermi momentum $p_F = 225$ MeV and binding energy $E_B=25$ MeV; the 
dotted and solid lines have been obtained using the LDA spectral function, 
 with and without inclusion of Pauli blocking, respectively.}
\end{center}
\end{figure}

The relative weight of different nuclear effects is illustrated in Fig.~\ref{total_elastic},
showing the total cross section of the process $\nu_e+^{16}O \to e^-+X$ as a 
function of neutrino energy.
The dot-dash line represents eight times the elementary cross section, i.e. the cross 
section for interactions of neutrinos on free neutrons. The dashed line is the result of 
the RFGM without inclusion of Pauli blocking (PB). The difference between the dot-dash and 
dashed lines, the latter being sizably smaller, results from the Fermi motion 
(with momenta not exceeding $p_F$) of the target nucleons.
The solid line corresponds to our results obtained from the LDA oxygen spectral function.
It clearly appears that the inclusion of NN correlations leads to a further reduction of the
cross section. This reduction is further enhanced if we include PB, as shown by the dotted line, 
since the available phase space for the knocked nucleon gets smaller, see Eq.~(\ref{pauli}).
Note that, as expected, the importance of PB decreases as neutrino energy increases.
For example, the difference between the solid and dotted lines turns out 
to be 10 \% and 6 \% at neutrino energy 0.5 GeV and 1.5 GeV, respectively.
It has to be mentioned that the prescription of Eq.(\ref{pauli}) can also be used to 
include PB in the RFGM. The size of the effect is the same as in the calculation
using the spectral function, the resulting cross section being suppressed by 
$\sim 5-10$ \% with respect to the dashed line of Fig. \ref{total_elastic}.
Overall, based on the results shown in Fig. \ref{total_elastic} one can safely state that 
a realistic treatment of neutrino-nucleus interactions must go beyond the RFMG. 
%(even if some complicated modifications, as that including 
%the nucleon potentials, are considered).

The same conclusions can be drawn for antineutrino interactions. Neglecting the isospin 
breaking
effect leading to the proton to neutron mass difference, the same Eqs.~(\ref{final_cs})-
(\ref{elem_cs}) can be applied. The only difference arises in the sign in front of 
the parity-violating
term of Eq.~(\ref{hadrdec}), $\varepsilon_{\mu\nu\alpha\beta}$, which should be replaced by 
$-\varepsilon_{\mu\nu\alpha\beta}$. Therefore, we do not show the corresponding figure.

\subsection{Comparison with other models}
\label{comp}

Over the past few years, many authors have investigated neutrino cross sections and the 
related nuclear effects (see, e.g., Ref. \cite{NUINTXX}).
In this section we compare our results with the existing literature
for the case of oxygen and carbon targets. It has to be kept in mind, however, that, while 
many of the available results correspond to $E_\nu \lsim 100$ MeV \cite{volpe}, 
the IA scheme adopted in our work is not expected to be accurate in this regime. 
Therefore, our comparison is mostly focused on the region of higher energies.

In Ref.\cite{nieves}, the CC quasi elastic cross section is obtained from the
self-energy of the gauge bosons $W^\pm$ in the nuclear medium, whose calculation
is carried out using the medium-modified nucleon propagator resulting from
a LDA implementation of the Fermi gas model. The effects of long range nuclear
correlation (RPA) are also included by means of an effective NN interaction of
the Landau-Migdal type. Coulomb distortion effects, accounting for the fact that the
charged lepton produced in neutrino interaction is moving in the Coulomb field of
the nucleus, are also taken into account. A collection of the resulting
$\sigma(^{16}O(\nu_e,\,e^-)X)$ is reported in Tab. IV of Ref. \cite{nieves}, in which
relativistic and non-relativistic nucleon kinematics, as well as the inclusion of
FSI and RPA effects are analyzed. 

\begin{figure}[b!]
\begin{center}
\epsfxsize11cm\epsffile{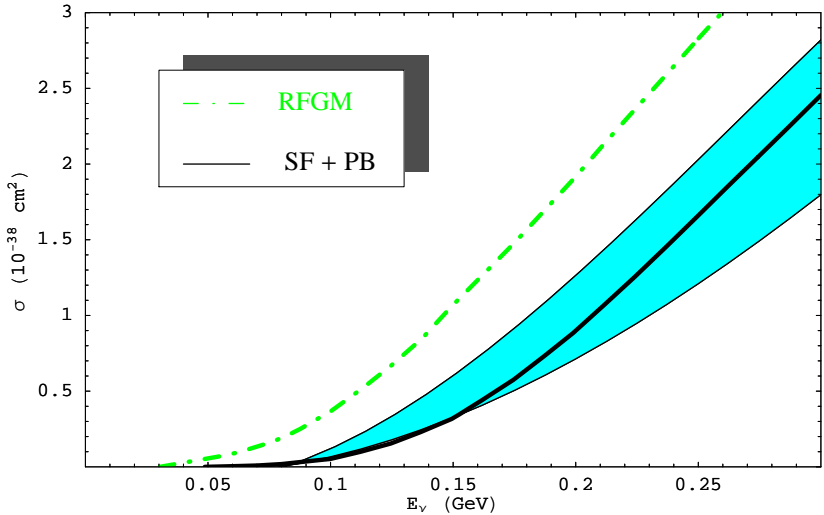}
\caption{\label{confnieves}\it
Comparison between our results for $\sigma(^{16}O(\nu_e,\,e^-)X)$ (solid line) and
those presented in \cite{nieves}; the shaded region represents the range of
cross sections predicted in Ref.\cite{nieves} (see text for details).
The RFGM results corresponding to Fermi momentum $p_F = 225$ MeV and $E_B=25$ MeV are
also shown.}
\end{center}
\end{figure}

In Fig.~\ref{confnieves} we compare our results to those of Ref.~\cite{nieves}.
The solid line corresponds to our calculation with the LDA spectral function and 
PB included; the shaded area is bounded by the minimum
and maximum values of the cross section of Ref.\cite{nieves}.
For reference, the RFGM prediction is also included.
It can be seen that the agreement between our results and those reported in 
Ref.\cite{nieves} is satisfactory. The small discrepancy at low neutrino energy,
below $\sim 0.15$ GeV, is likely to be ascribed to the inadequacy of the IA
to describe small momentum transfer processes. It is apparent thet the RFGM
largely overestimates the cross section.

\begin{figure}[h!]
\begin{center}
\begin{tabular}{c}
\epsfxsize8.5cm\epsffile{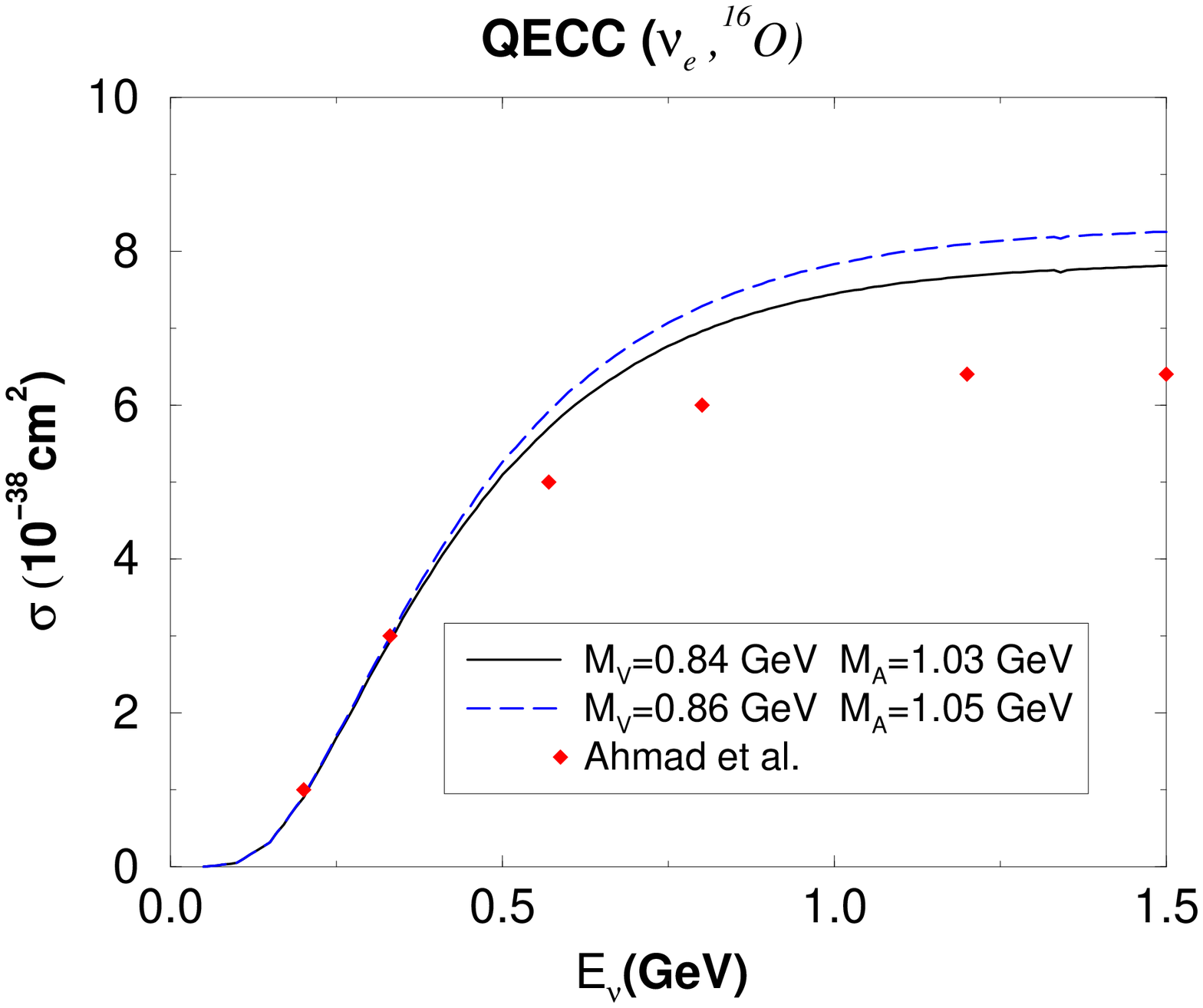} \\
\epsfxsize8.5cm\epsffile{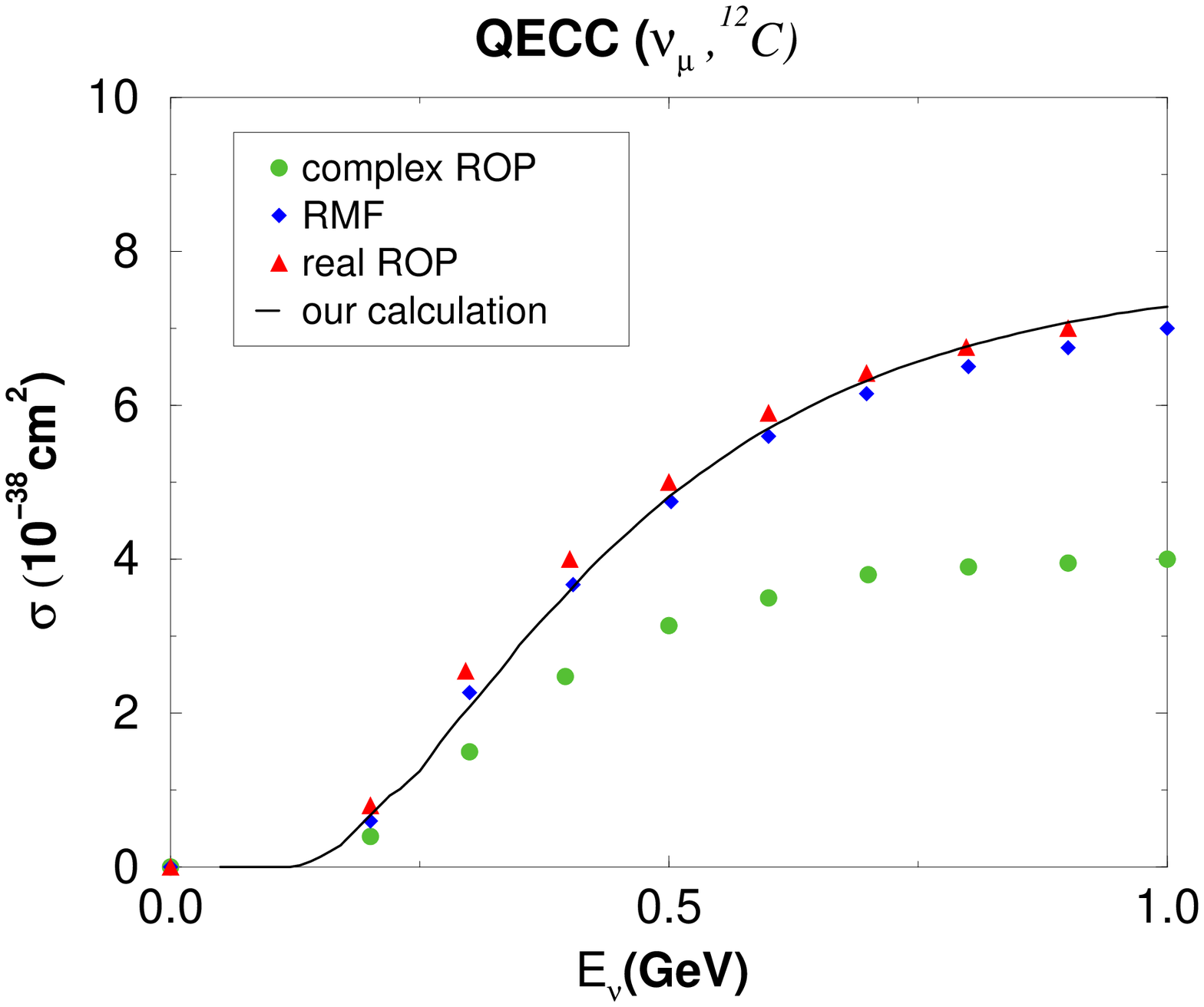}
\vspace{-0.4cm}
\end{tabular}
\caption{\it 
Top: comparison between $\sigma(^{16}O(\nu_e,\,e^-)X)$ resulting from our calculations
(solid and dashed lines) and that of Ref.\protect\cite{ahmad}. Bottom: 
comparison of $\sigma(^{16}O(\nu_\mu,\,\mu^-)X)$ resulting from our calculations
and those obtained in Ref.\protect\cite{maieron} using different models (see text for
details).}
\label{ahmad-maieron}
\end{center}
\end{figure}

At higher energies, we compare our results with those of Refs.\cite{ahmad} and \cite{maieron}.
In the first paper, $\sigma(^{16}O(\nu_e,\,e^-)X)$ is
calculated using the same formalism as described in \cite{nieves}. The
corresponding results are presented in the top panel of Fig.~\ref{ahmad-maieron}.

In order to pin down the source of the sizable disagreement between our results and those of 
Ref.\cite{ahmad}, we have tested the sensitivity of the cross section to the vector and 
axial masses using two different sets of values: $M_V=0.86$ and $M_A=1.05$, as employed by 
Ahmad {\it et al.} \cite{ahmad}, and $M_V=0.84$ and $M_A=1.03$.
It clearly appears that the discrepancies at $E_\nu > 0.5$ GeV cannot be removed by a different
choice of $M_V$ and $M_A$, and should rather be ascribed to the different descriptions
of nuclear dynamics.

A much better agreement is observed in the bottom panel of Fig.~\ref{ahmad-maieron}, in
which we compare our results for $\sigma(^{16}O(\nu_\mu,\,\mu^-)X)$, to those obtained in 
Ref.\cite{maieron} using different models. The approach of Maieron {\it et al.}, based 
on a IA scheme similar to the one adopted in our work, uses Relativistic Shell-Model wave
functions to describe bound nucleon states and two different prescriptions to include 
FSI effects. The final state of the knocked out nucleon is described using i) distorted waves 
obtained solving the Dirac equation with a phenomenological relativistic complex optical potential (ROP)
 or ii) the continuum solutions of the same Dirac equation used for the initial bound 
states (Relativistic Mean Field (RMF) approach.

The diamonds in the bottom panel of Fig.~\ref{ahmad-maieron}
show the RMF results of Ref. \cite{maieron}, whereas 
triangles-up and circles have been obtained by the same authors including the real 
part only and the full ROP (real + imaginary part), respectively.
The most stricking feature is the large effect of the imaginary part of the ROP, 
leading to a suppression of the total cross section of more than 40 \%.
This effect can be easily understood, as the imaginary optical potential produces a
loss of flux in the one particle-one hole channel. However, it contradicts the well 
known result stating that, in the kinematical regime in which incoherent
scattering dominates, {\it the integrated inclusive cross section is not affected by FSI}
\cite{horikawa}. In fact, it has long been realized that, unless the reappearence of the missing 
flux in the two particle-two hole channel is properly taken into account, the prescription 
employed in Ref. \cite{maieron} leads to a violation of the non energy weighted sum rule 
of the nuclear response \cite{horikawa}. On the other hand, 
Fig.~\ref{ahmad-maieron} shows that the cross sections obtained by Maieron {\it et al.} neglecting 
FSI effects or using only the real part of the ROP, whose effect is very small, are in 
excellent agreement with our results.

%%%%%%%%%%%%%%%%%%%%%%%%%%%%%%%%%%%%%%%%%%%%%%%%%%%%%%%%%%%%%%%%%%%%%%%%%%%%%%%%%%%%%%%
\section{Inelastic cross sections}
\label{resonance}

One pion production is the second process which contributes to the total cross 
section in the kinematical regime discussed in this paper;
it proceeds mainly through resonance production, the leading contribution coming from
excitation of the $\Delta$, i.e. through the 
the processes 
\be
\nu_\ell + n \to \ell + \Delta^+ \ \ \ , \ \ \ \nu + p \to \ell + \Delta^{++} \ .
\ee
In this section, we calculate the inclusive 
resonance production cross section applying the same formalism used for 
quasi-elastic interactions, in which nuclear effects in the initial state are described
by the spectral function. Obviously, PB has
no impact on the $\Delta$ produced at the weak interaction vertex, although it 
may affect the phase space available to the nucleon appearing in the aftermath 
of its decay. It has to kept in mind, however, that the inclusive 
cross section is only sensitive to processes taking place within a distance 
$\sim 1/|{\bf q}|$ of the primary vertex, so that at large enough $|{\bf q}|$
it is unlikely to be affected by two-step reaction mechanisms.

\subsection{$\Delta$ resonance production}

To account for $\Delta$ production, we can apply 
Eqs.~(\ref{final_cs})-(\ref{contraction}) with only minor changes.
Unlike the quasi-elastic case, the structure functions depend now on two variables, namely
$q^2$ and $W^2$, where $W^2$ is the squared hadronic invariant mass. The
energy conserving $\delta$-function is replaced by the Breit-Wigner factor
\be
\label{brfactor}
\frac{M_R\,\Gamma_R}{\pi}\, \frac{1}{(W^2-M_R^2)^2+M_R^2\,\Gamma_R^2} \ ,
\ee
where $M_R$ is the resonance mass and $\Gamma_R$ its
 decay width, and an additional integration in the variable $W$ must be carried out.
For the dependence of the structure functions on form factors for $\Delta^{++}$ production, 
we closely follow Ref. \cite{lalapas}\footnote{Notice that, compared to Eq.~(\ref{hadrdec}),
the parametrization of the hadronic tensor in \cite{lalapas} 
involves an additional factor $-1/2$ in the coefficient of $W_3$.} and use isospin symmetry 
to relate $\Delta^{++}$ and $\Delta^{+}$ form factors through
\be
\langle \Delta^{++}|J_\mu^A|p\rangle=\sqrt{3}\,
\langle \Delta^{+}|J_\mu^A|n\rangle \ .
\ee

It has to be mentioned that, in principle, the form factors entering the calculation 
of the cross section could be extracted from proton and
deuteron data. However, electron scattering studies show that 
extrapolations of the existing phenomenological models to the region of low $Q^2$ leads to 
sizably underestimate the nuclear cross sections in the $\Delta$ production 
region \cite{Benhar:2005dj,PRL06}.

\begin{figure}[h!]
\begin{center}
\epsfxsize9cm\epsffile{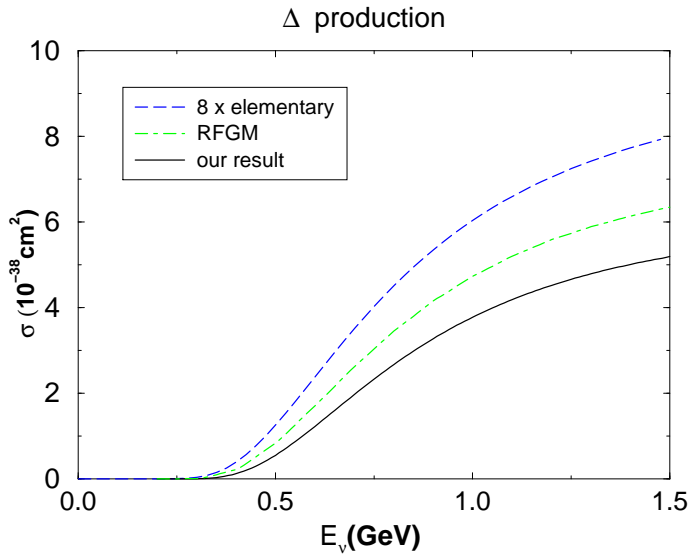}
\caption{\label{deltaprod}\it
CC $\Delta$ production via $^{16}O(\nu_e,e^-)$ interactions. The solid line shows the
results obtained using the LDA spectral function; the dashed and dot-dash curves are
the elementary cross section and the RFGM prediction (with $p_F = 225$ MeV and $E_B=25$ MeV), 
respectively.}
\end{center}
\end{figure}

Our results for CC $\Delta$ production via $^{16}O(\nu_e,e^-)$ interactions are shown
in Fig.~\ref{deltaprod}, and compared with the elementary cross
section and the RFGM prediction. 

As in the elastic case, nuclear effects strongly reduce the cross section with
respect to the free space value; moreover, the RFGM appears to
overestimate $\Delta$ production, an effect that has been already observed by 
different authors (see, e.g., \cite{nakamura2,sato}).

\begin{figure}[h!]
\begin{center}
\epsfxsize9cm\epsffile{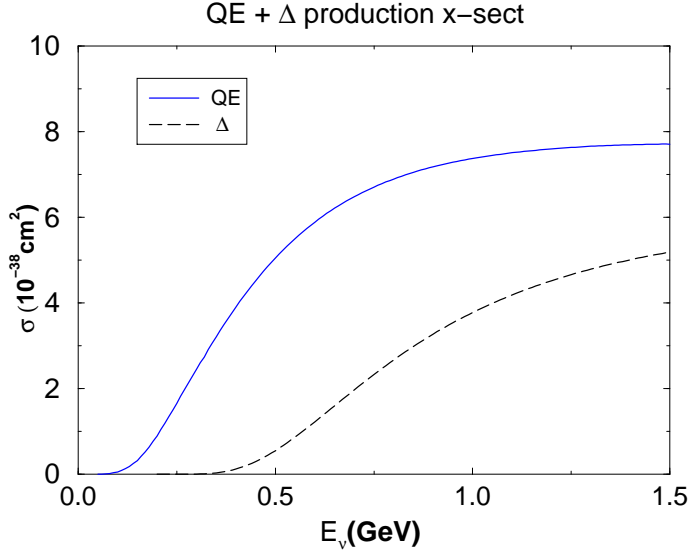}
\caption{\label{CCplusD}\it
Comparison between the CC quasi-elastic (solid line) and $\Delta$ production (dashed line) 
contributions to the total cross section of the process $^{16}O(\nu_e,e^-)$.}
\end{center}
\end{figure}

Fig. \ref{CCplusD} illustrates the relative weight of the quasi-elastic
and $\Delta$ production channels in the total inclusive $^{16}O(\nu_e,e^-)$ 
cross section.

\begin{figure}[h!]
\begin{center}
\epsfxsize9cm\epsffile{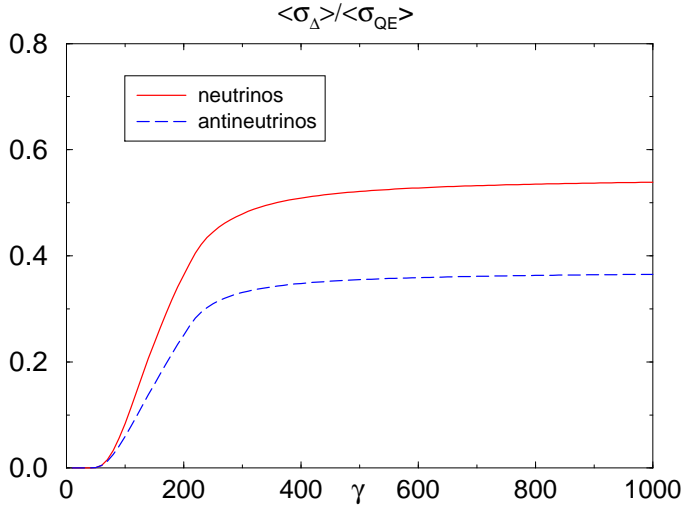}
\caption{\label{ratiogamma}\it
Ratio $ \langle \sigma_{\Delta} \rangle / \langle \sigma_{QE} \rangle $ as a function of the boost factor
$\gamma$ at $\beta$-Beams. The solid (dashed) line refers to neutrino
(antineutrino).}
\end{center}
\end{figure}

Above $E_\nu\sim 0.5$ GeV, the $\Delta$ production contribution becomes
important; in particular, the ratio of the areas under the curves is of the order of 40 \%, 
thus showing the importance of including such a contribution in
realistic estimates of the cross sections. 
To gauge the impact of $\Delta$ production on the measurements at $\beta$-Beams, we have computed
the averaged cross section
\bea
\label{ratiosigma}
\langle \sigma_i(\gamma) \rangle & = &\frac{\int dE_\nu\,\Phi_{\nu_e}(E_\nu,\gamma)
\,\sigma_i(E_\nu)}
{\int dE_\nu\,\Phi_{\nu_e}(E_\nu,\gamma)} \qquad i=QE,\Delta
\eea
for the fluxes reported in Ref.\cite{Burguet-Castell:2003vv}, 
and constructed the ratio $ \langle \sigma_{\Delta} \rangle / \langle \sigma_{QE} \rangle $ as a 
function of the boost factor $\gamma$. The results are shown in 
Fig.~\ref{ratiogamma}. 

It clearly appears that
the ratio of $ \langle \sigma \rangle$'s becomes larger and larger as the
$\gamma$ factor increases, until it becomes quite stable at $\gamma>600$ reaching the values
0.52 and 0.37 for neutrino and antineutrino, respectively. 
This implies that the $\Delta$ production contribution to the total cross section is about
$1/2$ or $1/3$ of the main contribution from CC quasi-elastic interactions.
The presence of the plateaux is due to the fact that at higher energies the
cross sections exhibit almost the same energy dependence.
Comparing with Fig.~\ref{fig:fluxes} we also understand the reason why the $\Delta$
production process contributes more to neutrino than to
antineutrino cross section. For fixed $\gamma$, the antineutrino spectra is peaked 
at lower energies than the neutrino one, in a region where the
contribution of the $\Delta$ is less important. In Appendix ~\ref{eventi} we
list separately the values of $\langle \sigma_{QE} \rangle$ and
$\langle \sigma_{\Delta} \rangle$ for the spectra of Fig.~\ref{fig:fluxes}.

\subsection{Contribution of higher resonances}

\begin{figure}[h!]
\begin{center}
\epsfxsize9cm\epsffile{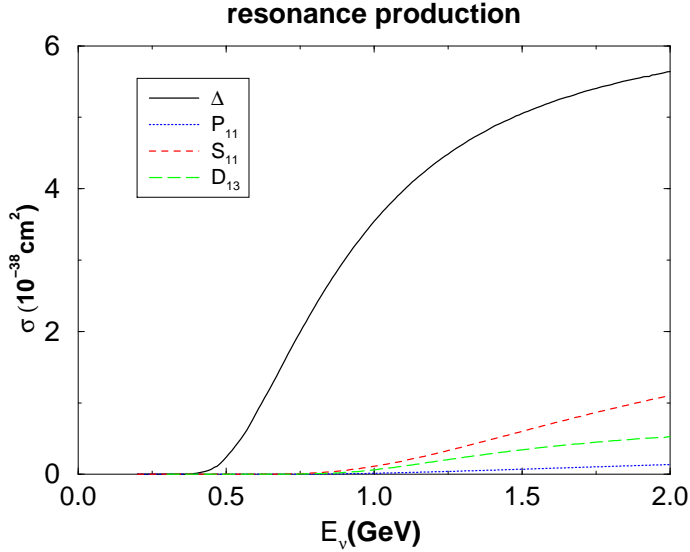}
\caption{\label{other}\it
CC resonance production via $^{16}O(\nu_e,e^-)$ interactions. The solid line shows 
the results for the $\Delta$ production, whereas the long dashed, dashed and dotted lines 
correspond to $D_{13}$, $S_{11}$ and $P_{11}$, respectively.}
\end{center}
\end{figure}

Beside the contribution of the $\Delta$ to the inelastic cross section, other
resonances may be important, particularly for neutrino energies larger than 
$\sim$ 2 GeV. This so-called {\it second resonance region} contains three isospin
1/2 states: $P_{11}(1440), D_{13}(1520), S_{11}(1535)$. New data on
electroproduction have enabled the authors of \cite{lalapas2} to estimate the
vector form factors of these resonances. On the other hand, estimates for the
axial couplings have been obtained from the decay rates of the resonances, 
exploiting the PCAC hypothesis. The strategy for computing their contributions to 
the cross section is the same used for $\Delta$ production, and we refer 
to \cite{lalapas2} for a compilation of the form factors.
Our results are shown in Fig.~\ref{other}.

As expected, even at the end of the energy spectra of Fig.~\ref{fig:fluxes},
the contribution of the second resonance turns out to be marginal. For example, at 
$E_\nu=1.5$ GeV, the ratio of the cross sections is 
$\sigma_\Delta:\sigma_{S_{11}}:\sigma_{D_{13}}:\sigma_{P_{11}}=1:0.12:0.06:0.02$.

\subsection{Comparison with other calculation and experimental data}

\begin{figure}[h!]
\begin{center}
\epsfxsize9cm\epsffile{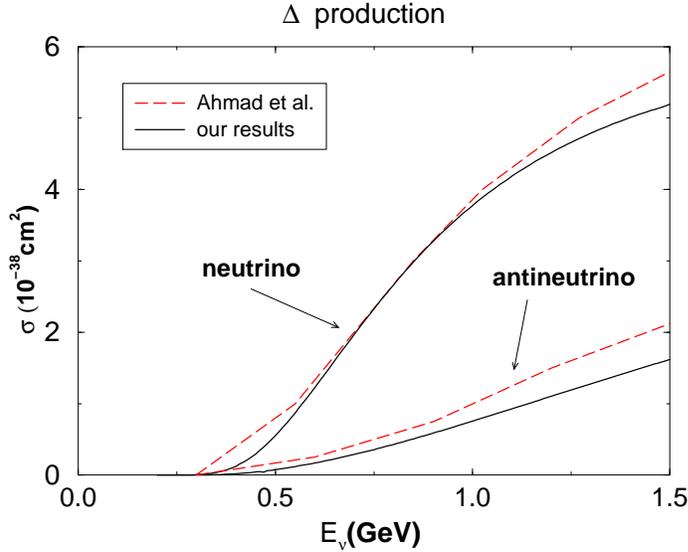}
\caption{\label{deltacomp}\it
Comparison of our cross sections (solid lines) for $^{16}O(\nu_e,e^-)$ and 
$^{16}O({\bar \nu}_e,e^+)$ to those of Ref. \cite{ahmad} (dashed lines).}
\end{center}
\end{figure}

As for the quasi-elastic case, many papers have been devoted to the subject 
of resonance production. However, most of them focus on exclusive channels (a subject 
only marginally addressed in the following) and consider targets other than 
oxygen or carbon \cite{ahmad2,leitner}. For the sake of completeness, in 
Fig.~\ref{deltacomp} we compare our calculation for $\Delta$ production in 
$^{16}O(\nu_e,e^-)$ interactions to the results of Ref.\cite{ahmad}. 
The agreement between the two calculations appears to be better than 
in the case of quasi-elastic scattering. This feature may be ascribed to the fact 
that the approach of Ref. \cite{ahmad} includes the effect of the $\Delta$ propagation 
through the nuclear medium, neglected in our calculation.

To end this section, we compare our calculations to the preliminary
results of the MiniBooNE collaboration for the cross section of the CC process
$\nu_\mu+^{12}~C \rightarrow \mu^-+\pi^++X$ \cite{Wascko:2006tx}. We should mention 
that a complete treatment of exclusive channels has not been carried out yet within our
formalism. In particular, we are able to estimate {\it incoherent} 
$\pi^+$ production (in which the target nucleus can be excited and/or broken up, 
 leading to the production of pions) whereas we have not taken into account the 
production channel $\nu+A\to \mu^-+ A+\pi^+$, 
in which the target nucleus is left in the ground state (coherent pion
production).

As the CC one $\pi^+$ production cross section is extracted from the measured ratio 
\bea
\label{ratiosigma2}
R=\frac{\sigma_{\pi^+}}{\sigma_{CCQE}}  
\eea
using a specific model for CC quasi-elastic cross section, we prefer to calculate 
$R$ from our model and then compare directly with Fig.~5 of \cite{Wascko:2006tx}.
We also include the contribution of the second resonance region. 

The following isospin decomposition can be applied to the pion production 
amplitudes $\mathcal{A}$:
\cite{leitner}
\begin{eqnarray}
\nn
\mathcal{A}(\nu_\ell + p \to \ell^- + p + \pi^+)&=& \mathcal{A}_3, \\
\mathcal{A}(\nu_\ell + n \to \ell^- + n + \pi^+)&=& 
\frac{1}{3}\mathcal{A}_3 + \frac{2 \sqrt{2}}{3} 
\mathcal{A}_1,  \\
\nn \mathcal{A}(\nu_\ell + n \to \ell^- + p + \pi^0)
&=& -\frac{\sqrt{2}}{3}\mathcal{A}_3 + \frac{2}{3} \mathcal{A}_1, \label{eq:ppnull} \ ,
\end{eqnarray}
where $\mathcal{A}_3$ is the amplitude for the isospin $3/2$ state of the 
$\pi N$ system, predominantly the $\Delta$, and $\mathcal{A}_1$ is 
the amplitude for the isospin $1/2$ state. 
From the above decomposition, we obtain the $\pi^+$ production cross section
\bea
\label{eq:nppl}
\sigma(\nu_\ell + p \to \pi^+)=\frac{10}{9}\,\sigma_{\Delta^{++}}+\frac{8}{9}\,
(b_1\,\sigma_{P_{11}}+b_2\,\sigma_{D_{13}}+b_3\,\sigma_{S_{11}}) \ ,
\eea
where the coefficients $b_i$ are branching ratios for $\pi^+$ production of
the resonances (which we assume to be equal to 0.7 for $P_{11}$, $D_{13}$ and $S_{11}$).

\begin{figure}[h!]
\begin{center}
\begin{tabular}{c}
\hspace{-0.3cm} \epsfxsize10cm\epsffile{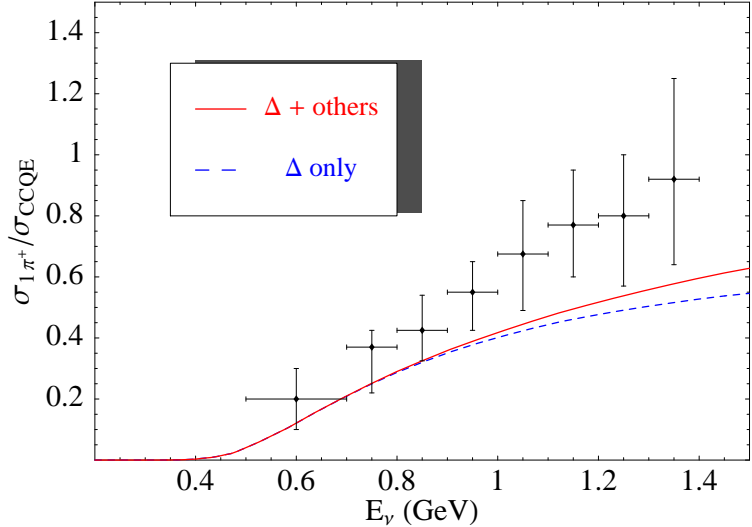}\\
\end{tabular}
\caption{\it \label{unpione}
Comparison between the preliminary MiniBooNE data for the ratio 
$R=\sigma_{\pi^+}/\sigma_{CCQE}$ in the process 
$\nu_\mu + ^{12}C \rightarrow \mu^- + \pi^+ + X$ and the results of our 
calculation. The dashed line refers 
to the $\Delta$ contribution only, whereas the solid line is obtained considering also
the contribution from the second resonance region (see Eq.~(\ref{eq:nppl})).}
\end{center}
\end{figure}

The results of the calculation are shown in Fig.~(\ref{unpione}) in which the
experimental points are taken from \cite{Wascko:2006tx}; the dashed line refers 
to the $\Delta$ contribution only, whereas the solid line is obtained using
Eq.~(\ref{eq:nppl}).

The figure shows a fair agreement between experiment and theory. However, at
energies above $E_\nu\sim 1$ GeV our curves are below the data. 
This may be ascribed to non-resonant pion production \cite{nakatalk} which has not been
taken into account in the calculation. Another $\sim 10$\% 
contribution could also come from coherent pion production \cite{Wascko:2006tx}.
%Moreover, the theoretical curves are
%strongly dependent on the value of the axial mass $M_A$.
A fully quantitative understanding of these effects deserves further studies.
%%%%%%%%%%%%%%%%%%%%%%%%%%%%%%%%%%%%%%%%%%%%%%%%%%%%%%%%%%%%%%%%%%%%%%%%%%%%%%%%%%%%%%%
\section{Conclusions}
In this paper we have investigated neutrino interactions with nuclei 
at energies up to $\sim 1$ GeV. In this energy region,
the main contributions to the total inclusive cross section come from CC
quasi-elastic and resonance production processes. To estimate
the cross sections, we employed a formalism based on IA, applicable
to both quasi-elastic and inelastic processes.

For the quasi-elastic $^{16}O(\nu_e,e^-)$ interactions, 
we compare various theoretical models, based on different descriptions of
nuclear dynamics. We find that the simple RFGM overestimates the 
cross sections, with respect to the results obtained using a realistic spectral
function to describe the strongly correlated bound nucleons. While dynamical
FSI play no role in the total inclusive cross section, statistical correlations,  
arising from the effect of Pauli principle, further decrease the cross 
section, as the phase space available to the knocked out nucleons 
is reduced. Therefore, our best estimate of the
inclusive total cross section is obtained using the LDA spectral function and 
including Pauli blocking. These results are in fair agreement with those 
discussed in the existing literature. 

Although the formalism we have employed can be readily applied to a variety of 
different targets, for which realistic spectral functions are available,
most of the available quasi-elastic neutrino data have been taken at 
low energies, where IA is not expected to provide accurate predictions. 
Higher energy data will soon be provided by K2K SciBar experiment \cite{private} 
and Miner$\nu$a \cite{manly}.

As for the inelastic cross sections, we focused on resonance production,
including the contribution of both the $\Delta$ and the second resonance region. 
The calculated inclusive resonance production cross sections show that, as in 
the elastic case, the simple RFGM appears to be unable to accurately describe
these processes. As expected, the CC resonanace production
to quasi-elastic cross section ratio becomes increasingly important at 
$E_\nu > 0.5$ GeV. At energies below $E_\nu \sim$2 GeV, the
contribution of $P_{11},D_{13}$ and $S_{11}$ to the total cross section 
turns out to be small, being $\sim$ 10\%. 

Even if we do not consider exclusive channels explicitely, we provide an 
estimate of the one $\pi^+$ production through resonance excitation. 
Comparison between the calculated ratio $R=\sigma_{\pi^+}/\sigma_{CCQE}$ to 
the preliminary results from the MiniBooNE collaboration shows a fair 
agreement between theory and data at $E_\nu \lsim 1$ GeV. On the other hand, at 
higher energies, a discrepancy emerges that may be removed once the contributions
of non-resonant and coherent pion production are properly added to our results.

%%%%%%%%%%%%%%%%%%%%%%%%%%%%%%%%%%%%%%%%%%%%%%%%%%%%%%%%%%%%%%%%%%%%%%%%%%%%%%%%%%%%%%%
\section*{Acknowledgments}
The authors are indebted to P. Lipari for a critical reading of the manuscript.
D.M. wishes to thank J.J. Gomez Cadenas, M. Sorel, G.P. Zeller and M. Wascko
for illuminating discussions on the status of the K2K and MiniBooNE cross section 
determination.
Many discussions with O. Lalakulich, M. Lusignoli and M. Sakuda are also 
gratefully acknowledged.
\appendix
\section{Integration limits}
\label{int_limit}
Let us analyze in detail the integration limits involved in the calculation of
the quasi-elastic and anelasic cross sections. The integration in momentum in
Eq.~(\ref{final_cs}) can be cast in the form 
\bea
\label{d3p}
d^3p = p^2 \,dp \,d\cos\gamma \,d\phi \ ,
\eea
where $\gamma$ is the angle between ${\bf p}$ and ${\bf q}$ and the integral
on the azimuthal angle $\phi$ can be readily done, yielding a factor $2\,\pi$. Denoting by
 $W$ the invariant mass of the final hadronic state produced in the
interaction, $\cos\gamma$ can be written in the form:
\bea
\nn
\cos\gamma=\frac{-W^2-|{\bf p}|^2-|{\bf q}|^2+(\tilde{\nu}+E_{\bf p})^2}
{2\,{\bf |{\bf p}|\,|{\bf q}|}} \ ,
\eea
or, equivalently,
\bea
\label{coseno}
\cos\gamma=\frac{s+M^2_{A-1}-W^2-2(\nu+M_A)\,E_{A-1}}
{2\,{\bf |{\bf p}|\,|{\bf q}|}} \ ,
\eea
where $M_{A-1}=M_A-m_N+E$, $s$ is the squared center of mass 
energy and $E_{A-1}^2=M^2_{A-1}+|{\bf p}|^2$.
The quantity $\cos\gamma$ has to satisfy the constraints $-1 \leq \cos\gamma \leq 1$, 
leading to the lower and upper bounds for $|{\bf p}|$:
\bea
|{\bf p}|^\pm=\frac{1}{2\,s}\,\left|\Lambda\,|{\bf q}| \pm
(\nu+M_A)\,\left[\Lambda^2-4\,s\,M^2_{A-1}\right]^{1/2}\right| \ ,
\eea
with $\Lambda=s+M^2_{A-1}-W^2$. 
The upper limit of the $E$ integration can be found requiring the argument of
the square root entering the definition of $|{\bf p}|^\pm$ to be
non-negative. This leads to $\Lambda \geq 2\,\sqrt{s}\,M_{A-1}$ and finally to
the bound
\be
E_{max}=\sqrt{s}-M_{A}-(W-m_N).
\ee
\subsection{Quasi-elastic case}
Up to now the integration limits we have found are completely general. The
quasi-elastic case is recovered once we put $W=m_N$.  
 Equation~(\ref{final_cs}) can be cast in a simpler form once
we use the energy conserving $\delta$-function to perform the integration on $\cos\gamma$ in
Eq.~(\ref{d3p}). We have to evaluate the Jacobian of the transformation
\bea
\nn \frac{\partial}{\partial \cos\gamma}\,(s - m_N^2) = 
 2 \ |{\bf p}|\,|{\bf q}| \ ,
\eea
leading to the final formula
\bea
\frac{d^2\sigma_{IA}}{d\Omega dE_\ell}=\frac{2\pi}{|{\bf q}|}\,
\int dp\,dE \,|{\bf p}| \,P({\bf p},E)\,
\frac{d^2\sigma_{\rm elem}}{d\Omega dE_\ell} \ .
\eea

\subsection{Resonance production}
For the case of resonance production, the variable $W$ is the invariant mass of
the resonance. It is more convenient to express Eq.~(\ref{d3p}) in terms of 
$|{\bf p}|$ and $W$ instead of $|{\bf p}|$ and $\cos\gamma$. It can be easily done
with the help of Eq.~(\ref{coseno}), implying 
\bea
d^3p = p^2 \,dp \,d\cos\gamma \,d\phi \Rightarrow 
\frac{2\pi}{|{\bf q}|}\,W\,dW\,|{\bf p}|\,dp \ .
\eea
The cross section for resonance production can then be written in the following
form:
\bea
\frac{d^2\sigma_{IA}}{d\Omega dE_\ell}=\frac{2\pi}{|{\bf q}|}\,
\int_{W_{th}}^{W_{max}}W\,dW \int_{p^-}^{p^+}|{\bf p}|\,dp\,
\int_0^{E_{max}}dE \,P({\bf p},E)\,
\frac{d^2\sigma_{\rm elem}}{d\Omega dE_\ell} \ ,
\eea
where $d^2\sigma_{\rm elem}/d\Omega dE_\ell$ has the same form as in
Eq.~(\ref{elem_cs}) but the hadronic tensor involves the Breit-Wigner
factor of Eq.~(\ref{brfactor}) and the structure functions are those specified
in Ref. \cite{lalapas}.

\section{Kinematical factors}
\label{a_coeff}
Here we report the kinematical factors appearing in the general formula for the
cross sections, after tensor contraction (Eq.~(\ref{elem_cs})):
\bea
\nn
A_1 & = &m_N^2\, (k\cdot k^{'})\\ \nn
A_2 &=& (k\cdot \tilde p)\,(k^{'}\cdot \tilde p)-\frac{A_1}{2}\\
\nn
A_3 &=&(k\cdot \tilde p)\,(k^{'}\cdot \tilde q)-(k\cdot \tilde q)\,(k^{'}\cdot \tilde p)
\\
\nn
A_4&=&(k\cdot \tilde q)\,(k^{'}\cdot \tilde q)-\frac{\tilde q^2}{2}\,\frac{A_1}{m_N^2}\\
%\nn
A_5&=&(k\cdot \tilde p)\,(k^{'}\cdot \tilde q)+(k^{'}\cdot \tilde p)\,(k\cdot
\tilde q)-(\tilde q\cdot \tilde p)\,\frac{A_1}{m_N^2} \ ,
\label{def:A}
\eea
where
\bea
k&=&(E_\nu,{\bf k}) \qquad k^{'}=(E_\ell,\bf k^{'}) \cr \nn \cr \nn
\tilde p&=&(E_{|\bf p|},{\bf p}) \qquad \tilde q=(\tilde \nu,\bf q) \ .
\nn 
\eea
Notice that, due to the simmetry of the problem, some of the scalar products
considerably simplify. In fact, using a reference system in which the vector 
${\bf q}$ is along the z-axis and scattering takes place in the $xz$-plane, 
one can express the scalar products in terms of momentum component in that plane
and, using the symmetry of the problem, neglect terms linear in $p_x$.
The evaluation of the remaining contributions is straightforward.

\begin{figure}[h!]
\begin{center}
\begin{tabular}{c}
\epsfxsize9cm\epsffile{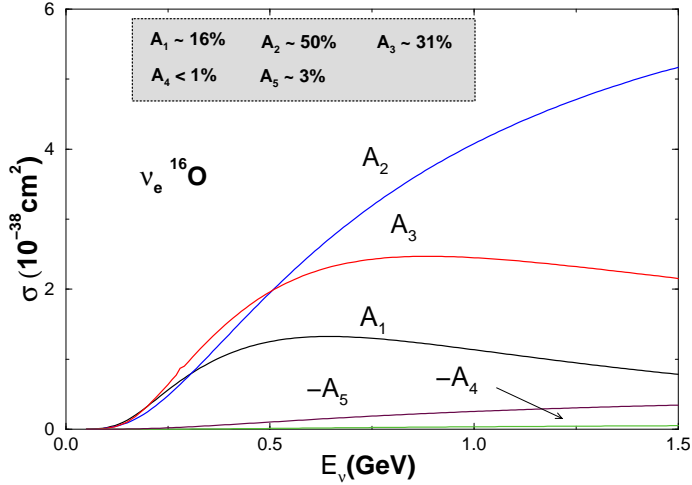}
\end{tabular}
\caption{\it \label{fig:relative}
Contributions of the different structure functions, defined in Eq.(\ref{def:A}), to the 
quasi-elastic process $\sigma(^{16}O(\nu_e,\,e^-)X)$. The values given in the shaded box
correspond to the areas under the different curves.}
\end{center}
\end{figure}

In the limit in which $\tilde q \to q$ and  $\tilde p \to p$, the coefficients $A_{2-5}$ 
reduce to kinematical factors of free elastic interactions. In particular, $A_{4,5}$ depend
quadratically on the lepton mass, thus vanishing in the limit of massless leptons. 
This is no
longer true in our formalism, in which, even in the limit of massless leptons, $A_{4,5}$ are
different from zero. The relative contribution of each term appearing in 
Eq.~(\ref{contraction}) is shown in Fig.~\ref{fig:relative} in which we plot the quasi-elastic
cross section for electron neutrinos on $^{16}O$.

While the main contributions come from $A_{1-3}$, 
non-vanishing $A_{4,5}$ give an additional 4\%  to the cross section . 
For heavier lepton masses, the contribution of $A_{4,5}$ is increasingly larger, as expected
for free nucleon interactions.

\section{Form factors for quasi-elastic scattering}
\label{formfactors}
The hadronic current appearing in the definition of the hadronic tensor in 
Eq.~(\ref{hadronictensor}) has the $V-A$ structure $J^\mu=J^\mu_V-J^\mu_A$. The most geenral
expressions of the vector and axial contributions read
\bea
\label{nuclcurr}
J^\mu_V&=&J^{\mu\,+}_V=\bar U(p^{\prime})\,\Gamma_V^\mu\,\tau^+\,U(p) \cr
\nn
J^\mu_A&=&J^{\mu\,+}_V=\bar U(p^{\prime})\,\Gamma_A^\mu\,\tau^+\,U(p) \ ,
\eea
where $p$ ($p^{\prime}$) are the momenta of the incoming (outgoing) nucleon and 
 $\Gamma_V^\mu$ and $\Gamma_A^\mu$ are given by:
\bea
\nn
\Gamma_V^\mu&=&\gamma^\mu\,F_1+i\,\sigma^{\mu\,\nu}\,q_\nu \frac{F_2}{2\,m_N}+
q^\mu \,F_S \\
\nn
\Gamma_A^\mu&=&\gamma^\mu\,\gamma^5\,F_A+i\,\gamma^5\,\sigma^{\mu\,\nu}\,q_\nu \,F_T+
q^\mu \,\gamma^5\,F_P
\eea

In the definition of $J^\mu_V$ and $J^\mu_A$, $\tau^+$ is the isospin raising
operator acting on the isodoublet $U=(n,p)^T$ and $F_1,F_2,F_S,F_A,F_P,F_T$ are
form factors depending on the squared four-momentum transfer $q^2$. The CVC hypothesis
allows one to relate the weak form factors to the eletromagnetic form factors and to
constrain $F_S (q^2)=0$. On the other hand, the PCAC hypothesis relates 
$F_P(q^2)$ and $F_A(q^2)$ in such a way that
$F_P(q^2)=2\,m_N\,F_A(q^2)/(m^2_\pi-q^2)$. Finally, as a {\it
second class} current is incompatible with the Standard Model of
electroweak interactions \cite{lee}, we set $F_T (q^2)=0$.

To express the structure functions in
terms of form factors we need to add the $VV$, $AV$ and $AA$ contributions
to the hadronic tensor (\ref{hadronictensor}) and equate the resulting sum 
to the decomposition (\ref{hadrdec}). Thus, we obtain:
\bea
\nn       W_1 &=& 2\,\left[ -\frac{q^2}{2}\,\left( F_1 + F_2 \right)^2 
          + \left( 2\,m_N^2 - \frac{q^2}{2}\right)\,F_A^2 \right]\cr
	  \nn \cr
	  \nn 
      W_2 &=& 4\,\left[F_1^2 -\left(\frac{q^2}{4\,m_N^2}\right)\,F_2^2 + 
      F_A^2 \right] \nn \cr \nn \cr
      W_3 &=& -4\,\left( F_1 + F_2 \right)\,F_A\nn \cr  \nn \cr 
      W_4 &=& -2\,\left[ F_1\,F_2 + \left( 2\,m_N^2 + \frac{q^2}{2} \right)\,
      \frac{F_2^2}{4\,m_N^2}+\frac{q^2}{2}\,F_P^2 - 2\,m_N\,F_P\,F_A \right]\nn
      \cr \nn \cr 
      W_5 &=& \frac{W_2}{2}\nn \ .
\eea
Defining the electric ($G_E$) and magnetic ($G_M$) form factors
\begin{equation}
      G_E = \frac{1}{\left( 1 - \frac{q^2}{M_V^2}\right)^2} \qquad 
      G_M = 4.71\,G_E  \ ,
\end{equation}
and $\tau = q^2/4\,m_N^2$, 
%\begin{equation}
%\tau = \frac{q^2}{4\,m_N^2} \ ,
%\end{equation}
we finally get
\bea
\nn
      F_1 &=& \frac{1}{ 1 - \tau }\,( G_E -\tau\,G_M )\qquad 
      F_2 = \frac{1}{ 1 - \tau }\,(-G_E + G_M)\cr \nn \cr \nn 
      F_A &=& -\frac{1.26}{\left( 1 - \frac{q^2}{M_A^2}\right)^2}\qquad 
      F_P = \frac{1.28}{\left( 1 -
      \frac{q^2}{0.14}\right)^2}\,\left(\frac{F_A}{1.27}\right) 
      \eea

\section{Averaged cross sections}
\label{eventi}
In Table~(\ref{tableeventi}) we show the results of $\bar \sigma$ for the quasi-elastic and
$\Delta$ production induced by neutrino and antineutrino, for the two
$\beta$-beams scenarios described in the paper.

\begin{table}[hbtp]
\begin{center}
\begin{tabular}{|c|c|c|c|} \hline
   & $\gamma$ &  $\bar \sigma_{QE}\, (10^{-38}$ cm$^2$) & 
   $\bar \sigma_{\Delta} \,(10^{-38}$ cm$^2$)  \\ \hline \hline
$\nu_e$ & 100 & 3.3 &  0.28  \\ \hline
$\bar \nu_e$ & 60 & 0.34 & 0.002\\ \hline \hline
$\nu_e$ & 250 & 6.55 & 2.9  \\ \hline
$\bar \nu_e$ & 150 & 1.35 & 0.21\\
\hline
\hline
\end{tabular}
\vspace*{.2in}
\end{center}
\caption{\it Average cross section $\bar \sigma$ (see Eq.(\ref{ratiosigma})) for quasi-elastic
scattering and $\Delta$ production induced by neutrino and antineutrino. The two different 
sets of $\gamma$ values correspond to the fluxes shown in Fig.~\ref{fig:fluxes}.}
\label{tableeventi}
\end{table}
%%%%%%%%%%%%%%%%%%%%%%%%%%%%%%%%%%%%%%%%%%%%%%%%%%%%%%%%%%%%%%%%%%%%%%%%%%%%%%%%%%%%%%%

\end{document}